# Development of AI-integrated infrastructure with biomedical device and mobile app for neonatal vital monitoring during and in between kangaroo care sessions

Saptarshi Purkayastha[1], Hrishikesh Bhagwat[1], Keerthika Sunchu[1], Orlando Hoilett[23], Eddy Odari[3], Reuben Thuo[3], Martin Wafula[4], Celia Kariuki[5], and Sherri Bucher[5]

*Abstract*—Premature infant mortality remains a critical challenge in low- and middle-income countries (LMICs), with continuous vital sign monitoring being essential for early detection of life-threatening conditions. This paper presents an integrated system combining NeoWarm, a novel biomedical device, with NeoRoo, a mobile application, and NeoSmartML, a machine learning infrastructure, to enable comprehensive vital sign monitoring during Kangaroo Mother Care (KMC). Our power-optimized device achieves 6-6.5 days of continuous operation on a single charge, while the mobile application implements an offline-first architecture with efficient data synchronization. The optical character recognition pipeline demonstrates promising accuracy (F1 scores 0.78-0.875) for automated vital sign extraction from existing NICU monitors. Experimental validation shows the system's feasibility for deployment in resource-constrained settings, though further optimization of heart rate and temperature detection, along with the risk classification foundation model is needed.

*Clinical relevance*: Wearable sensor technology combined with digital health tools embedded with machine learning decision support can equip and empower nurses and doctors with accurate real-time clinical information in settings where health worker shortages are pernicious, and routinely result in heavy patient volume.

## I. INTRODUCTION

Premature newborns face significant mortality risks in low- and middle-income countries (LMICs), with over 2.4 million annual deaths primarily from preventable causes [1]. Continuous monitoring of vital signs - respiratory rate (RR), heart rate (HR), blood oxygen ($SpO_2$), and body temperature - is crucial for early detection of lifethreatening conditions [2]. These parameters provide critical cardiorespiratory stability and thermoregulation indicators, essential for premature infant survival [2] [3].

Traditional contact-based monitoring methods using adhesive electrodes and sensors can cause skin irritation and damage in premature infants due to their extremely fragile skin [4]. Additionally, frequent repositioning of sensors increases infection risks in their immunocompromised state [5]. These challenges have motivated research into contactless monitoring approaches. There are some contactless approaches, such as the EarlySense under mattress sensors [6] and, remote-sensing cameras [7] [8]. However, their availability, complexity, and lack of infrastructural support make them experimental technologies that are difficult to scale [9].

While vital sign monitors are present in many NICUs, they often lack standardized data output formats and interfaces for automated data collection, making it difficult to systematically track and analyze patient data during critical interventions like STS [10]. Manual documentation by healthcare workers is prone to errors and inconsistencies, particularly in resource-constrained settings where staff must simultaneously monitor multiple infants [11].

Kangaroo Mother Care (KMC), particularly skin-to-skin (STS) contact, is an evidence-based neonatal intervention that improves premature infant outcomes through improved body temperature regulation, antibody generation, and maternal-infant bonding [12]. However, implementing STS in resource-constrained settings faces significant challenges: understaffed facilities struggle with continuous monitoring, and inadequate training limits proper vital sign tracking during STS sessions [13] [14].

To address these challenges, we developed NeoWarm [15] [16], a patented biomedical device [17], and NeoRoo [18], a companion mobile application. This integrated system enables accurate vital sign monitoring [19] during STS while providing automated session tracking and alerts. Our approach enables continuous, automated documentation of both session timing and physiological parameters without requiring expensive monitor upgrades or complex integration protocols. The infrastructure encompasses three key innovations:

1) The NeoWarm biomedical device implements lowpower continuous vital sign streaming [20] [21] to NeoRoo, ensuring extended battery life while maintaining data accuracy. This enables prolonged monitoring without disrupting STS sessions or requiring frequent charging - crucial for LMIC settings with unreliable power supply.

[1] Luddy School of Informatics, Computing and Engineering, Indiana University Indianapolis, IN, USA.
[2] Department of Biomedical Engineering, University of Cincinnati, OH, USA
[3] Jomo Kenyatta University of Agriculture and Technology, Nairobi, Kenya
[4] Mama Lucy Kibaki Hospital, Nairobi, Kenya
[5] Fairbanks School of Public Health, Indiana University Indianapolis, IN, USA

2) We developed a novel data collection system using ESP32 cameras to capture vital sign data from diverse NICU monitors. This approach anonymously collects labeled vital signs with risk categorization, creating a comprehensive dataset representing varied premature infant conditions across different monitoring systems.
3) We trained a foundation model on this collected data and optimized it for deployment on widely deployed cheap smartphones. This model operates offline within the NeoRoo mobile app, analyzing streaming data from NeoWarm to generate real-time risk alerts based on vital sign patterns.

This paper focuses on validating these design choices through experimental evaluation before human trials for the NeoWarm device, as seen in Figure 1. We demonstrate the system's feasibility in terms of power efficiency, data collection accuracy, and model performance on resourceconstrained smartphones. Our approach aims to provide accessible, reliable vital sign monitoring during KMC, potentially reducing premature infant mortality in resourcelimited settings, including those which suffer from pernicious nursing shortages [22].

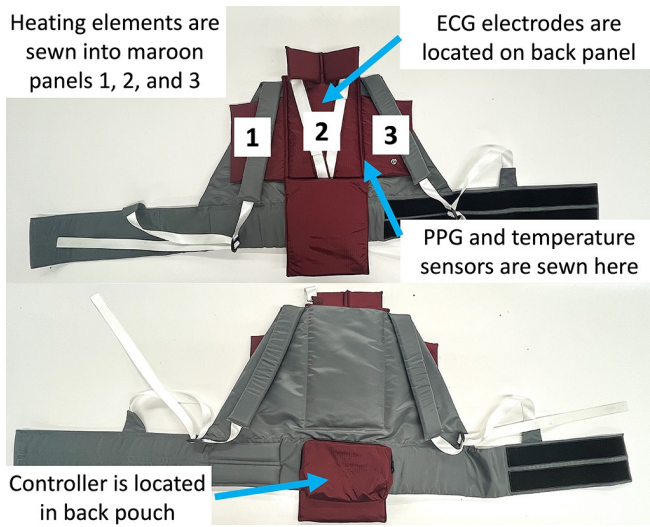

Fig. 1. The NeoWarm Device

## II. SYSTEM ARCHITECTURE - NEOWARM + NEOROO + ML

### A. Hardware Layer - NeoWarm Device

The NeoWarm device employs a nRF52840 multiprotocol Bluetooth 5.4 SoC featuring an ARM Cortex-M4 core operating at 64 MHz along with a floating point unit (FPU). The processor has an on-chip flash memory of size 1 MB and 256 KB of RAM memory. Power supply through a high-efficiency DC-DC converter and an on-chip advanced adaptive power management system achieves exceptionally low energy consumption [23]. The processor has an onchip access control list peripheral designed to assign and enforce access restriction schemes for different regions of the on-chip flash memory [24] and an advanced encryption standard (AES) CCM peripheral to ensure data security while maintaining real-time performance for sensor data acquisition and processing.

The sensor array comprises three medical-grade components optimized for neonatal monitoring. The MAX30102 pulse oximeter and heart-rate sensor employs dualwavelength photoplethysmography with 100 Hz sampling, providing SpO2 accuracy of ±1.39% and pulse rate measurements within ±2.04 beats per minute (bpm) [25]. The MLX90614 medical-grade infrared (IR) thermometer delivers continuous temperature monitoring with ±0.2 °C accuracy and 0.02 °C resolution, crucial for detecting subtle thermal variations in premature infants. Electrocardiogram (ECG) electrodes interface through an AD8232 analog frontend, featuring dynamically configurable filter settings to remove motion artifacts and isolate R-waves for determining HR. The AD8232 also includes an integrated right leg drive (RLD) for enhanced signal quality in high-noise environments.

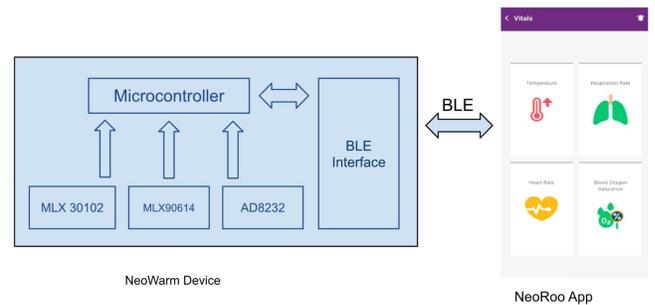

Fig. 2. Functional components of the NeoWarm device

Power management is implemented through a sophisticated hierarchical system that maintains reliable operation while minimizing energy consumption. The ultra-low-power architecture achieves <10 $\mu$A sleep current through selective peripheral shutdown and CPU sleep modes, with rapid wakeup capabilities (<50 $\mu$s) for critical measurements. Motionbased adaptive sampling dynamically adjusts sensor acquisition rates, reducing power consumption during periods of low activity while maintaining high-frequency sampling during movement or clinical interventions. BLE transmission is optimized through data batching and compression, transmitting aggregated vital signs at a 1 Hz default rate with configurable burst modes for critical events. This comprehensive power management strategy enables continuous operation with vital signs data sent over Bluetooth Low Energy for over 72 hours on a single 1 Ah lithium polymer (LiPo) battery (with dimensions 4.0 x 2.8 x 0.4 cm, weighing approximately 18 grams i.e. less than 3% of the weight of a premature baby) while maintaining medical-grade monitoring accuracy.

*B. Mobile Application Layer - NeoRoo*

The NeoRoo mobile application, with 2 customized user interfaces (one for parents, the other for healthcare providers) is built using Google's cross-platform framework, Flutter [26]. The screens of the mobile application are designed to be responsive across different screen sizes and orientations ensuring a smooth user experience on a wide range of Android devices, which are predominant in LMIC settings [27]. The mobile application architecture follows the Business Logic Component (BLoC) pattern [28], an event-based state update mechanism, enabling efficient user interface (UI) updates while preventing redundant rerendering of UI components. The app maintains 60 frames per second (FPS) performance while processing concurrent vital sign streams from up to 20 NeoWarm devices, and subsequently 20 mother-infant dyads within a KMC ward.

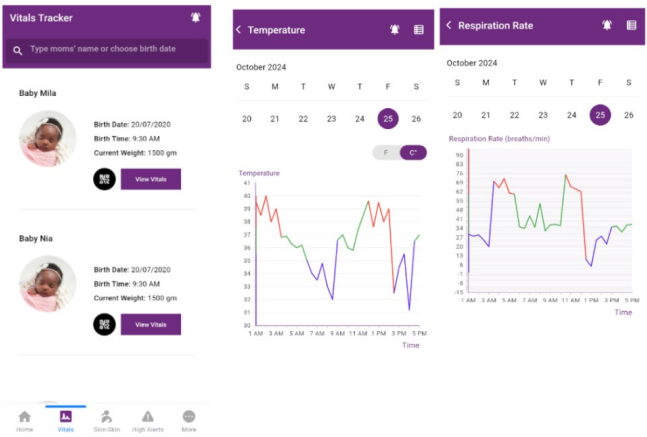

Fig. 3. The NeoRoo app allows us to view vital signs of multiple neonates in real-time with feature to sync collected data with DHIS2

We use Hive [29] database, with AES-256 encryption for securely storing time-series vital signs data locally along with clinical annotations and device metadata. The offline-first architecture, designed for LMICs with unreliable internet access, syncs with DHIS2 [30] servers using a timestamp-based conflict resolution strategy and sequential FIFO processing, maintaining integrity of data. Data transfer overhead was reduced through delta synchronization, which only transmits changed data values since the last successful sync, and Deflate compression algorithm with optimized LZ77 sliding window for time-series medical data. Leveraging the DHIS2 file system API, the app also features a learning resource section to educate parents on neonatal care; healthcare providers can access a library of materials, and push targeted evidence-based educational messaging to parental caregivers at different points within the clinical care pipeline. This is designed to improve trust, integration of care, and task-sharing among key stakeholders (i.e., family members and healthcare providers) in a neonate's life.

The connection interval is dynamically adjusted by a connection manager based on received signal strength and priority, with intervals ranging from 7.5 ms during critical operations to 400 ms during standard operations. Automatic reconnection handling utilizes an exponential backoff strategy with device-specific connection profiles, while the batched data transmission system implements a variable-length encoding scheme that achieves 40-54% data compression [31] while maintaining sub-second latency for critical vital sign updates.

*C. Infrastructural Layer - NeoSmartML*

We propose NeoSmartML, a system for automated data acquisition, processing, and machine learning model development to health conditions of the premature neonates. NeoSmartML collects vital signs data by sampling NICU monitor images, combines it with annotations provided by healthcare workers, and trains predictive models. To extract key vital signs—including heart rate (HR), blood oxygen saturation ($SpO_2$), and respiratory rate (RR)—we developed an optical character recognition (OCR)-based pipeline, similar to prior work [32] but improved to handle variations in monitor location and color. This approach addresses the challenge of data interoperability in healthcare settings where monitors often lack standardized data output interfaces [33]. The trained models can be integrated into the NeoRoo mobile application to enable real-time outcome prediction at the point of care.

The system implementation utilizes an ESP32-CAM module for image acquisition, selected for its optimal balance of processing capabilities and energy efficiency [34]. The module was programmed with custom firmware to capture images at 1 Hz sampling frequency, aligning with clinical requirements for continuous vital sign monitoring in neonatal care. For system validation, we curated a dataset of 45 images from publicly available resources, displayed on monitors, and captured using the ESP32-CAM to simulate diverse NICU environments. This methodology ensures robustness across various monitor designs, display resolutions, and ambient lighting conditions.

The OCR pipeline was implemented using EasyOCR [35], a deep learning-based text recognition library. Before text extraction, the images underwent preprocessing steps to enhance recognition accuracy [36]. Contrast enhancement was applied to improve text visibility under varying lighting conditions, and noise reduction techniques were used to eliminate background artifacts and screen glare. Geometric corrections, such as image rotation and perspective alignment, ensured that text regions were properly aligned for effective OCR analysis. Bounding boxes around detected text regions were expanded horizontally and vertically to capture numerical values adjacent to vital sign labels such as HR, $SpO_2$, and RR.

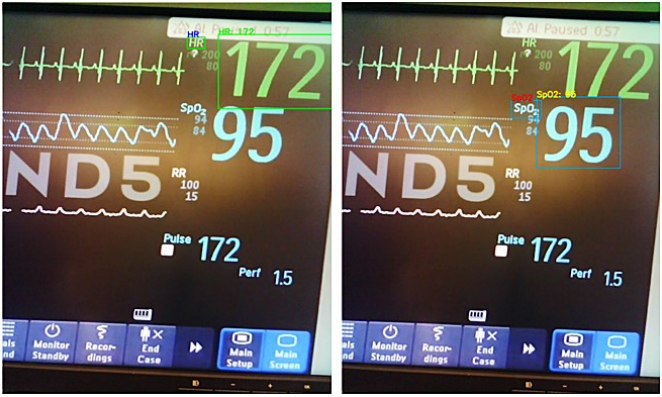

Fig. 4. Neonatal vital sign displays: HR and SpO₂

Our implementation initializes the OCR with English language configuration and GPU acceleration for enhanced processing efficiency. The system employs a hierarchical detection approach, first identifying vital sign labels (e.g., "HR", "ECG") as anchor points, then localizing corresponding numerical values through spatial relationship analysis. This methodology builds upon established techniques in medical display text extraction [37]

Within the expanded region, the pipeline searched for numeric text. Each detected number was evaluated to ensure it fell within the defined search area. The bounding box size of each numeric value was calculated, and the value with the largest bounding box was selected. This heuristic was adopted because, on most NICU monitors, the actual vital sign value (e.g., HR) is displayed prominently with a larger font size than the smaller numbers representing ranges or historical data near the label. This approach ensured that the most visually prominent numerical value closest to the HR label was correctly identified. A similar methodology was employed for SpO₂ and RR, to address the unique display patterns and layouts associated with these vital signs. The pipeline leveraged parallel processing using Python's concurrent futures library to optimize efficiency, enabling multiple images to be processed simultaneously.

This infrastructure demonstrates the feasibility of automated vital sign extraction across diverse NICU monitoring environments, particularly valuable in resource-constrained settings where manual documentation remains prevalent [22]. The system's integration into NeoSmartML provides a scalable foundation for continuous neonatal monitoring, contributing to improved healthcare accessibility in lowresource contexts.

*D. Neural Network Architecture*

The Streaming Multimodal Transformer (SMT) architecture processes three distinct data streams through a novel attention mechanism optimized for real-time neonatal monitoring. The continuous vital sign stream (temperature, heart rate, SpO₂, respiratory rate) is processed at 1 Hz through a sliding window approach with a 5-minute context, while static features (maternal age, previous complications, birth weight) and semi-static features (weekly blood markers, feeding patterns) are encoded through separate embedding layers.

The Streaming Crossmodal Transformer (SCT) forms the core processing unit, utilizing a 3x3 kernel with four attention heads operating in 64-dimensional space. The relative positional encoding scheme employs sinusoidal basis functions with learnable frequency components, enabling the model to capture both fine-grained temporal patterns in vital signs and longer-term clinical trends. The attention mechanism implements a sparse computation strategy that reduces computational complexity from $O(n^2)$ to $O(n \log n)$ while maintaining medical-grade accuracy through selective attention to clinically significant temporal regions.

The Feature Fusion Module integrates multimodal inputs through a hierarchical attention mechanism that dynamically weights feature importance based on clinical context. The classification head produces risk stratification (low/moderate/high) with calibrated probability estimates, utilizing focal loss with dynamic class weighting to address the inherent imbalance in critical event frequency.

*E. System Integration - Data Flow, Alert Generation and Security Implementation*

The integrated system implements a multi-tiered architecture optimized for reliable operation in resource-constrained environments, building upon the previously described NeoWarm, NeoRoo, and NeoSmartML components. Real-time vital signs from NeoWarm are processed through an efficient ring buffer implementation, enabling concurrent read/write operations without blocking while ensuring data integrity. The system employs priority-based queuing, processing critical vitals at 1 Hz while handling static features through configurable batch intervals. This approach complements the power optimization strategies of NeoWarm while supporting the mobile app's performance. There are no known regulatory requirements for the frequency of vitals capture, but we use 1Hz as a norm across various ICU monitors.

The alert generation system combines traditional thresholdbased monitoring (normal/abnormal values) with machine learning predictions from NeoSmartML. The NeoSmartML approach can use time-series and historical values that traditional threshold-based monitoring missing out on. Clinical thresholds are customizable per patient category (extreme preterm, very preterm, etc.), while a Bayesian framework incorporates sensor reliability metrics and historical patterns to minimize false alarms. Alert aggregation employs temporal clustering with dynamic windows, addressing the challenge of alarm fatigue highlighted in previous sections while maintaining clinical effectiveness.

The security architecture implements comprehensive protection meeting healthcare standards (HIPAA/GDPR)

through multiple layers. Data transmission between NeoWarm and NeoRoo uses ChaCha20-Poly1305 encryption, while stored data employs AES-256-GCM with perfect forward secrecy through ephemeral keys. The system leverages hardware security modules when available, with software-based white-box cryptography as a fallback for devices without dedicated secure elements. Access control combines biometric verification with role-based permissions using JWT tokens that support offline operation, crucial for LMIC environments with intermittent connectivity.

### III. METHODS - Validation Experiments

*A. Evaluation of NeoWarm Device*

The validation methodology for NeoWarm focuses on power consumption and sensor accuracy. Sensor validation employs medical-grade reference devices: a Masimo Radical-7 for $SpO_2$/pulse rate comparison, GE Healthcare CARESCAPE for ECG verification, and a Braun Thermoscan Pro 6000 for temperature accuracy. Bluetooth LowEnergy (BLE) performance testing uses radio frequency (RF) chambers to measure transmission reliability under varying signal conditions. To assess the power consumption of the device, we power the device using a 1 Ah LiPo battery connected to a MAX17048 fuel gauge. The setup was configured for update intervals of 1, 2, and 5 s for two hours each. Current consumption was measured for each of these update intervals. All the sensors were powered through the output voltage pins of the microcontroller.

*B. Automated Testing of the NeoRoo App*

The NeoRoo app was developed based on the Test Driven Development philosophy, which includes automated tests for app logic, offline functionality, and database synchronization. The sync mechanism undergoes validation using simulated network conditions with varying latencies (50-2000 ms) and packet loss (0-30%). Concurrent client testing employs automated scripts simulating multiple devices performing simultaneous data modifications. BLE stream processing validation utilizes recorded vital sign datasets played back through multiple NeoWarm devices simultaneously, measuring packet loss and latency under conditions. The DHIS2 integration testing framework verifies data consistency across the distributed system using checksums and versioning metadata.

*C. Evaluation of NeoSmartML*

NeoSmartML validation methodology encompasses OCR accuracy, model training efficiency, and inference performance. OCR testing utilizes a dataset of 10,000 monitor screenshots from five different vendor displays under varying lighting conditions (100-1000 lux) and viewing angles (±30 degrees). The ESP32 firmware undergoes stress testing with continuous capture over 72-hour periods, monitoring memory usage and system stability. Model training validation implements k-fold cross-validation with stratified sampling across different risk categories. Inference performance testing measures latency and memory usage across various Android devices (Android 8.0-13.0) with different CPU/RAM configurations.

Integration testing of the complete system employs automated test harnesses simulating real-world usage patterns. The test environment includes multiple NeoWarm devices streaming simultaneously to several NeoRoo instances, with varying network conditions and user interactions. Resource utilization monitoring tracks CPU, memory, and battery usage across all system components. Security testing includes penetration testing of the REST APIs, BLE communication protocol analysis, and verification of data encryption at rest and in transit. The testing infrastructure logs detailed metrics for each component, enabling analysis of system behavior under various operational conditions.

### IV. RESULTS

*A. NeoWarm Device Power Consumption*

We conducted a comprehensive power consumption analysis of the NeoWarm device across multiple operational scenarios to evaluate its energy efficiency and battery life implications. The analysis was performed over onehour intervals for each test condition to ensure stable measurements and account for any periodic variations in power draw.

Initially, the device was evaluated in its advertising mode, where it continuously broadcasted its capabilities over Bluetooth Low Energy (BLE) radio frequency, consuming an average current of 12.79 mA. This baseline measurement represents the device's power requirements during the discovery phase before establishing a connection with a mobile device, as seen in Fig. 5.

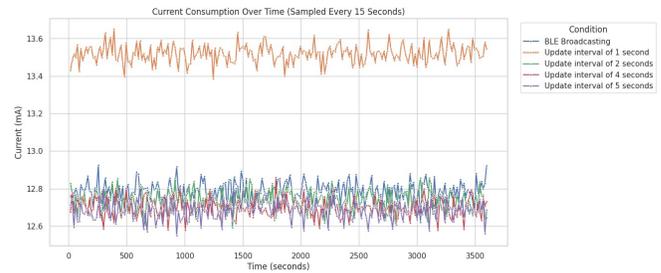

Fig. 5. Current consumption observed over 1 hour in each scenario

Subsequently, we analyzed the device's power consumption during active data transmission with varying update intervals. With the shortest update interval of 1 second, representing the highest data transmission frequency, the device consumed 13.52 mA. This modest 5.7% increase from the advertising mode demonstrates the efficient implementation of our BLE stack. As the update interval was increased to 2, 4, and 5 seconds, the current consumption progressively decreased to 12.74 mA, 12.70 mA, and 12.69 mA, respectively, as shown in Table I.

TABLE I
CURRENT CONSUMPTION OF NEOWARM IN DIFFERENT SCENARIOS

| Condition | Current |
|---|---|
| BLE Broadcasting | 12.79 mA |
| Update interval of 1 second | 13.52 mA |
| Update interval of 2 seconds | 12.74 mA |
| Update interval of 4 seconds | 12.70 mA |
| Update interval of 5 seconds | 12.69 mA |

The minimal variation in power consumption across different update intervals (maximum deviation of 0.83 mA) indicates effective power management implementation. The actual NeoWarm product might be deployed with a standard 3.7V, 2000mAh (36 g) LiPo battery, suggesting an operational life of approximately 148-157 hours (6-6.5 days) on a single charge, making the device suitable for extended monitoring sessions. The relatively flat power consumption profile across different update intervals provides flexibility in adjusting data transmission frequency based on clinical requirements without significantly impacting battery life. These results demonstrate that our power optimization strategies, including efficient BLE implementation and careful hardware component selection, successfully achieve the balance between reliable vital sign monitoring and extended battery life necessary for continuous neonatal care in resource-constrained settings.

*B. OCR Results*

To evaluate the performance of the OCR pipeline for extracting neonatal vital signs, we compared the values identified by the pipeline with manually annotated ground truth values for HR, SpO$_2$, and RR. The F1 score, which combines precision, recall, and accuracy, was used as an evaluation metric. The results are presented in Table II:

TABLE II
F1 SCORES AND ACCURACY FOR OCR PERFORMANCE

| Metric | F1 Score | Accuracy |
|---|---|---|
| Heart Rate (HR) | 0.78 | 0.64 |
| Oxygen Saturation (SpO$_2$) | 0.86 | 0.75 |
| Respiratory Rate (RR) | 0.875 | 0.77 |

The pipeline achieved its highest F1 score and accuracy for RR (0.875 and 0.77, respectively), followed by SpO$_2$ (0.86 and 0.75) and HR (0.78 and 0.64). The lower accuracy for HR is attributed to challenges such as overlapping text, small font sizes, and varying monitor layouts, which affected the OCR pipeline's ability to extract values reliably.

## V. DISCUSSION

Our integrated NeoWarm-NeoRoo-NeoSmartML system demonstrates promising capabilities for neonatal vital sign monitoring in resource-constrained settings, while also highlighting areas for continued development. The power consumption results of the NeoWarm device, showing operational lifetimes of 6-6.5 days on a single charge, compare favorably with existing wearable monitoring solutions that typically require daily charging [20][21]. This extended battery life is particularly crucial for LMIC settings where reliable power access remains a challenge, as noted by Imam et al. [22] in their systematic review of nursing care in acute care hospitals.

The OCR pipeline's performance shows varying levels of accuracy across different vital signs (F1 scores ranging from 0.78 to 0.875), with respiratory rate achieving the highest accuracy. These results align with similar work by Jeon et al. [32], who reported comparable challenges in extracting heart rate values from monitor displays. The lower accuracy for heart rate extraction (F1 score of 0.78) may be attributed to the greater variability in heart rate display formats and positions across different monitor manufacturers, a challenge also noted by Rampuria et al. [33] in their analysis of ICU monitor data extraction.

The system's offline-first architecture and delta synchronization approach addresses key connectivity challenges identified by Geldsetzer et al. [27] in their review of healthcare applications in LMICs. Our implementation of the Deflate compression algorithm with optimized LZ77 sliding window demonstrates effective data handling while maintaining clinical utility, though the achieved compression ratios suggest potential for further optimization. This is particularly relevant given the bandwidth constraints commonly encountered in LMIC healthcare settings.

The integration of vital sign monitoring with KMC support represents a novel approach to addressing the challenges highlighted by Kinshella et al. [14] regarding facility-based KMC implementation. Our system's ability to provide automated session tracking while maintaining continuous vital sign monitoring could help address the staffing constraints that often limit KMC implementation in resource-constrained settings. However, the current accuracy levels of the OCR system suggest that human verification of critical values remains necessary, aligning with the hybrid monitoring approach advocated by Mahato et al. [9].

These findings need to be considered within the context of certain limitations. The OCR system validation was conducted using a relatively small dataset of 45 images, and real-world performance may vary under different lighting conditions and monitor configurations. Additionally, while our power consumption analysis demonstrates extended battery life, long-term reliability testing under actual clinical conditions will be necessary to validate these results. Future work should focus on expanding the training dataset for the OCR system, optimizing the machine learning models for lower-end mobile devices, and conducting comprehensive clinical validation studies.

## VI. CONCLUSIONS

This paper presents an integrated system combining hardware, mobile application, and machine learning components to enable comprehensive vital sign monitoring during Kangaroo Mother Care. Our experimental validation demonstrates the feasibility of extended battery operation, efficient data handling through compression and synchronization, and automated vital sign extraction through OCR, though with varying levels of accuracy across different parameters. While further optimization and clinical validation are needed, the overall architecture shows promise for addressing critical monitoring needs in low-resource healthcare environments, potentially improving access to continuous vital sign monitoring during KMC sessions where manual documentation remains challenging.

ACKNOWLEDGMENT

This work is being conducted under the auspices of the NeoInnovate Collaborative Consortium, headquartered at IU Indianapolis and encompassing a multidisciplinary collaborative effort across various US and international institutions. We gratefully acknowledge support of the Fairbanks School of Public Health – Jomo Kenyatta University of Agriculture and Technology Partnership Working group, including: Dr. Patrick Mburugu; Dr. Musa Mohamed; Dr. Steven Ger Nyanjom; Dr. Serah Kaggia; Ms. Hellen Mwai. We thank Sew Valley (Cincinnati, OH) for textile manufacturing assistance (Fig. 1).